\documentclass[10pt,preprint]{aastex}

\shorttitle{XMM-Newton Observations of Radio Pulsars} \shortauthors{Gil, Haberl, Melikidze et al.}

\begin{document}
\author{J.~Gil\altaffilmark{1}, F.~Haberl\altaffilmark{2}, G.~Melikidze\altaffilmark{1,3}, U.~Geppert\altaffilmark{4}, B.~Zhang\altaffilmark{5}
and G.~Melikidze Jr.\altaffilmark{1}}
\altaffiltext{1}{J. Kepler Institute of Astronomy, University of Zielona G\'ora, Poland}
\altaffiltext{2}{Max Planck Institute for Extraterrestial Physics, Garching, Germany}
\altaffiltext{3}{E. Kharadze Georgian National Astrophysical Observatory, Tbilisi, Georgia}
\altaffiltext{4}{German Aerospace Center, Institute for Space Systems, Berlin, Germany}
\altaffiltext{5}{Department of Physics, University of Nevada, Las Vegas, USA}

\title{XMM-Newton Observations of Radio Pulsars B0834+06 and B0826-34 and
Implications for Pulsar Inner Accelerator}

\def\be{\begin{equation}}
\def\ee{\end{equation}}
\def\lesssim{\raisebox{-0.3ex}{\mbox{$\stackrel{<}{_\sim} \,$}}}
\def\gtrsim{\raisebox{-0.3ex}{\mbox{$\stackrel{>}{_\sim} \,$}}}
\def\degsp{\hbox{$^{\circ}$}}
\def\EB{\hbox{${\rm {\bf \Delta E} \times {\bf B_s}}$}}
\def\xmm{XMM-Newton}

\newcommand{\ergs}[1]{$\times 10^{#1}$ erg s$^{-1}$}
\newcommand{\hcm}[1]{$\times 10^{#1}$ cm$^{-2}$}
\newcommand{\nh}{N$_{\rm H}$}
\newcommand{\lbol}{\hbox{$L_b$}}
\newcommand{\fbb}{\hbox{F$_{\rm bb}$}}
\newcommand{\fpl}{\hbox{F$_{\rm pl}$}}
\newcommand{\ct}{cts s$^{-1}$}



\label{firstpage}

\begin{abstract}
We report the X-ray observations of two radio pulsars with drifting
subpulses: B0834$+$06 and B0826$-$34 using \xmm\. PSR B0834$+$06 was
detected with a total of 70 counts from the three EPIC instruments over 50
ks exposure time. Its spectrum was best described as that of a blackbody
(BB) with temperature $T_s=(2.0^{+2.0}_{-0.9}) \times 10^6$~K and
bolometric luminosity of $L_b=(8.6^{+14.2}_{-4.4}) \times 10^{28}$
erg~s$^{-1}$. As it is typical in pulsars with  BB thermal components in
their X-ray spectra, the hot spot surface area is much smaller than that
of the canonical polar cap, implying a non-dipolar surface magnetic field
much stronger than the dipolar component derived from the pulsar spin-down
(in this case about 50 times smaller and stronger, respectively). The
second pulsar PSR B0826$-$34 was not detected over 50 ks exposure time,
giving an upper limit for the bolometric luminosity $L_b \leq 1.4 \times
10^{29}$ erg~s$^{-1}$. We use these data as well as the radio emission
data concerned with drifting subpulses to test the Partially Screened Gap
(PSG) model of the inner accelerator in pulsars. This model predicts a
simple and very intuitive relationship between the polar cap thermal X-ray
luminosity ($L_b$) and the ``carousel'' period ($P_4$) for drifting
subpulses detected in the radio band. The PSG model has been previously
successfully confronted with four radio pulsars whose $L_b$ and $P_4$ were
both measured: PSR B0943$+$10, PSR B1133$+$16, PSR B0656$+$14, and PSR
B0628$-$28. The \xmm\ X-ray data of PSR B0834$+$16 reported here are also
in agreement with the model prediction, and the upper limit derived from
the PSR B0826$-$34 observation does not contradict with such a prediction.
We also include two other pulsars PSR B1929$+$10 and B1055$-52$ whose
$L_b$ and/or $P_4$ data became available just recently. These pulsars also
follow the prediction of the PSG model. The clear prediction of the PSG
model is now supported by all pulsars whose $L_b$ and $P_4$ are measured
and/or estimated.

\end{abstract}


\section{Introduction}

More than forty years after the discovery of radio pulsars, the mechanism by which they emit coherent radio beams is still
not fully understood. Also, many properties of this radiation remain a mystery, especially the phenomenon of drifting
subpulses. This puzzling phenomenon was widely regarded as a powerful tool for the investigation of the pulsar radiation
mechanism. Recently, this phenomenon received renewed attention, mostly owing to the newly developed techniques for the
analysis of the pulsar radio emission fluctuations (Edwards \& Stappers 2002,2003; ES02,ES03 henceforth). Using these
techniques, Weltevrede et al. (2006 a,b; W06a,b henceforth) presented results of the systematic, unbiased search for the
drifting subpulses and/or phase stationary intensity modulations in single pulses of a large sample of pulsars. They found
that the fraction of pulsars showing evidence of drifting subpulses is at least 60~\% and concluded that the conditions for
the drifting mechanism to work cannot be very different from the emission mechanism of radio pulsars.

It is therefore likely that the drifting subpulse phenomenon originates
from the so-called inner acceleration region right above the polar cap,
which powers the pulsar radiation. In the classical model of Ruderman \&
Sutherland (1975; RS75 henceforth) the subpulse-associated spark filaments
of plasma circulate in the pure Vacuum Gap (VG hereafter) around the
magnetic axis due to well known drift of plasma with non-corotational
charge density (see Appendix A for more details). There are few
periodicities characteristic for this model, called also the pulsar
carousel model: the primary period $P_3$ which can be measured as a
distance between the observed subpulse drift bands, the secondary period
(apparent when drifting is aliased; Gil \& Sendyk 2003 for detailed
description), and the tertiary period $P_4$ (called also the carousel
time\footnote{designated as $\hat P_3$ in RS75. Although this symbol is
still in use, we advocate to replace it by $P_4$.}, as it is the time
interval after which the gap plasma completes one full circulation around
the magnetic pole). The carousel model is widely regarded as a natural and
qualitative explanation of the drifting subpulse phenomenon. However, its
original version published by Ruderman \& Sutherland (1975; RS75
hereafter) predicts too high a drifting rate of the sparks around the
polar cap, as compared with the observations of drifting subpulses (e.g.
Deshpande \& Rankin 1999, 2001; DR99,DR01 henceforth), and too high a
heating rate of the polar cap (PC henceforth) surface due to the
spark-associated back-flow bombardment, as compared with X-ray
observations (e.g. Zhang et al. 2000). Another difficulty of the RS75
model is that recent calculations strongly suggest that the surface
binding energy of both ions and electrons are too low to allow the
development of a vacuum gap. Indeed, when the surface magnetic field is
purely dipolar, then the gap can develop only in magnetars and several
highest B-field pulsars (Medin \& Lai 2007). Another type of inner
accelerator model, named space-charge-limited flow (SCLF, Arons \&
Scharlemann 1979; Harding \& Muslimov 1998), has been discussed in the
literature, which assumes that both ions and electrons can be freely
striped off the neutron star surface. Although this approximation is valid
for most pulsars assuming a pure dipolar field at the polar cap region, a
stronger, multipole magnetic field near the polar cap region (which is
needed to make a large number of radio pulsars above the radio emission
death line, Ruderman \& Sutherland 1975; Zhang et al. 2000) would
introduce a non-negligible binding energy of ions/electrons (Medin \& Lai
2007), which renders the SCLF approximation no longer valid. Another
difficulty of the steady-state SCLF model widely discussed in the
literature is that it does not predict the existence of any ``sparks''
that could give rise to the drifting sub-pulses. So, in our opinion, it is
not an attractive inner accelerator model to interpret pulsar radio
emission.

Motivated by these shortcomings of the otherwise attractive VG model Gil,
Melikidze \& Geppert (2003; G03 henceforth) developed further the idea of
the inner acceleration region above the polar cap by including the partial
screening caused by the thermionic flow of ions from the PC surface heated
by sparks. We call this kind of the inner acceleration region a "Partially
Screened Gap" (PSG hereafter). The PSG is thermally self-regulated in such
a way that the surface temperature is always close to but slightly lower
(less than 1 percent) than the critical temperature at which the maximum
co-rotational ion outflow occurs and the gap is fully screened (see
Appendix and/or G03 for more details). Moreover, if the surface
temperature was even few percent lower than the critical temperature,
there would be a pure vacuum gap, with all the problems discussed above.
Since the actual potential drop in the PSG is much lower than that of the
pure VG model (RS75), the intrinsic drift rate and PC heating rate are
compatible with measurements of $P_4$ and $L_b$, respectively.

The PSG model can be tested if two observational quantities are known: (i) the circulational period $P_4$ for drifting subpulses observed in
radio-emission and (ii) the X-ray bolometric luminosity $L_b$ of thermal BB radiation from the hot polar cap (see Appendix A). Radio pulsars were
targeted since beginning of X-ray astronomy for various scientific reasons. Zhang, Sanwal \& Pavlov (2005; Z05 henceforth) were the first who made an
attempt to resolve the mystery of drifting subpulses in radio pulsars by observing them in X-rays. They proposed to detect thermal X-ray photons from
the PC heated by sparks of plasma likely to be associated with drifting subpulses observed in radio band. Their choice was the best studied drifting
subpulse pulsar B0943+10. Using \xmm\  X-ray observatory they detected a weak source coincident with the target pulsar. Due to very small number of
counts detected, no unambiguous spectrum could be obtained. However, they were able to fit the BB model to the data, although a power law model was
acceptable as well. Within a BB model they inferred a bolometric luminosity $L_b \sim 5\times 10^{28}$ erg/s emitted from the hot spot (few MK) with
a surface area much smaller (about 60 times) than the conventional polar cap area as defined by the bundle of last closed dipolar field lines. This
radio pulsar was well studied by DR99, who described the number of sparks and the circulation time $P_4=37.4 P$ needed for them to complete one full
revolution around the pole (where $P$ is the basic pulsar period). These properties as well could not be accounted for by the conventional theory,
and some radical modification of RS75 model was required. It appears that PSG model not only resolves all the problems of the RS75 model, but also
offers a clean prediction that can be used to test theories of the inner pulsar accelerator.

\section{Previous work}

Gil, Melikidze \& Zhang (2006b; Paper I henceforth) reanalyzed the B0943+10 case within the PSG model. They derived a very
useful formula directly connecting the drifting rate of plasma sparks (measured by the circulation period $P_4$) and the
polar cap heating rate by the back-flow spark bombardment (measured by the bolometric thermal luminosity $L_b$). By assuming
that both the measured quantities are determined by the same value of electric field in the PSG, they obtained a simple
formula relating the so-called efficiency of thermal radiation from the hot polar cap with the circulation time
\begin{equation}
\frac{L_b}{\dot E} = 0.63\left(\frac{P_4}{P}\right)^{-2},
\end{equation}
where $\dot E$ is the pulsar spin-down (see eq.~[A3] with $I_{45}=\alpha=1$ in
Appendix~A). PSR B0943+10 with its data specified in Table 1, fitted this observational curve quite well (Fig. 1). When one
observable parameter in equation~(1) is known ($L_b$ or $P_4$), the other one can be predicted without any free parameters.
In Paper I we included B1133+16, the twin pulsar to B0943+10 (at least in the sense of the kinematical properties; see Table
1). In this second case we speculated that the long periodicity of about $30P$ revealed by a number of authors (e.g.
W06a,b), is actually the circulational period $P_4 \sim 30P$. This claim was recently confirmed by sophisticated data
analysis of Herfindal \& Rankin (2007; HR07 henceforth), although these authors admitted that they did not believe our
prediction of $P_4$ value before their own analysis. The X-rays from B1133+16 were detected by Kargaltsev, Pavlov and
Garmire (2006) using Chandra X-ray observatory, who found that their properties were similar to those of the twin pulsar
B0943+10. Because of the small number of counts detected, obtaining an unique spectrum was not possible, like in the case of
PSR B0943$+$10 (ZSP05) . However, the BB model was acceptable and gave the bolometric luminosity $L_b \sim 3\times 10^{28}$
erg/s emitted from the hot (few MK) and very small polar cap (again much smaller (about 100 times) than the canonical one).
As one can see in Figure~1, with the inferred values of $P_4$ and $L_b$ the pulsar B1133+16 nicely clusters with its twin
pulsar along the critical curve expressed by equation~(1). Note that filled circle represents our prediction and asterisk
represents the estimate of $P_4$ by HR07.

Encouraged by the observational confirmation of our prediction of $P_4$ in B1133+16, we applied the same method to two other
pulsars for which the measurements or estimates of thermal bolometric luminosity were available (Gil, Melikidze \& Zhang
2007; Paper II henceforth). One of the famous Three Musketeers, PSR B0656+14, in which thermal X-rays from small hot polar
cap were clearly detected by De Luca, Caraveo, Mereghetti, et al. (2005; DL05 hereafter), was an obvious choice. The BB
thermal luminosity $L_b \sim 5.7 \times 10^{31}$ ergs/s (Table 1) inserted into equation~(1), returned the predicted value
of $P_4=20.6~P$. Amazingly, Weltevrede et al. (2006c; W06c henceforth) reported the long-period fluctuation spectral feature
$(20\pm 1)P$ associated with the quasi-periodic amplitude modulation of erratic and strong radio emission detected from this
pulsar. Thus, it was tempting to interpreted this period as the circulation time $P_4$. With this value of $P_4$ and $L_b$
shown above, the pulsar B0656+14 fits the equation~(1) quite well (Figure~1). Although the drifting subpulses were not
apparent in this case, the erratic radio emission reported by W06c was similar to the so-called Q-mode in PSR B0943+10
(showing clearly drifting subpulses in the organized B-mode). The low frequency feature in the fluctuation spectra,
identical to that of the B-mode, was found by Rankin \& Suleymanova (2006; see their Fig.~6). Asgekar \& Deshpande
(2001;~AD01 hereafter) also detected this feature in the 35-MHz observations of PSR B0943+10(see their Figs.1 and 2). This
simply means that the carousel plasma drift is maintained in both regular drifting and erratic (with no drifting subpulses)
pulsar emission modes. This is a property of plasma and magnetic field interaction in the gap rather than the structure of
this plasma. However, drifting subpulses can be clearly observed only if the gap plasma has some lateral structure,
localized sparking discharges for instance.

For the second of the Three Musketeers, PSR B1055-52, we have just found
an evidence of a low frequency feature $f \sim 0.042 c/P$ (Biggs 1990; B90
henceforth), which can be interpreted as the carousel periodicity $P_4/P
\sim 22$. Using this interpretation, which was very fruitful in several
other cases discussed above and below, we examine thermal X-ray radiation
from the small hot spot detected in this pulsar and attempt to test our
PSG model in section 4.4. The third Musketeer (Geminga) is radio quiet, so
although it shows thermal BB X-ray emission from the small hot spot, it is
not useful for our analysis.

Another pulsar that we could examine using our method of inferring values of $P_4$ from intensity modulation spectra was PSR
B0628-28. As indicated in Table 1, it was detected in X-rays by Tepedelenlio\v{g}lu \& \"Ogelman (2005; T\"O05 henceforth),
using Chandra and \xmm\  observatories. This was an exceptional pulsar (called an overluminous one by Becker et al. 2005)
with efficiency much larger than that of typical pulsars (Becker \& Tr\"umper 1998). For thermal BB component alone
$L_b/\dot E
\sim 1.9 \times 10^{-2}$ (Table 1). This value inserted to equation~(1) gives the predicted value of $P_4 \sim (6 \pm 1) P$.
Interestingly, W06c reported for this pulsar a relatively short periodicity of $(7 \pm 1) P$ (Table~1). If this periodicity
is interpreted as the circulation time $P_4$, then this is pulsar is not exceptional at all. It lies on the theoretical
curve (eq.~[1]) in Figure~1 at exactly the right place. PSR B0628-28 is just another (fourth) pulsar satisfying the
predictions of equation~(1), which relates the efficiency of thermal X-ray radiation from a hot polar cap to the
circulational periodicity associated with drifting subpulses observed in radio emission.

In order to expand the sample of pulsars that have both $L_b$ and $P_4$
measured/estimated, we recently launched an observational campaign using
the \xmm\ Observatory. We targeted at two old pulsars that had $P_4$
measurements but had no X-ray observations before, and we report on the
results of these observations in this paper. The two pulsars, PSR B0826-34
and PSR B0834$+$06 were observed during the \xmm\  Cycles AO-5 and AO-6,
respectively. Simultaneous radio monitoring was also performed and we will
report on these observations in the separate paper. PSR B0826-34 was not
detected and we have derived an upper limit for its thermal luminosity. We
clearly detected PSR B0834$+$06, whose spectrum is best modelled by a BB
radiation from a small hot spot. We interpret this as due to PC heating by
the back-flow bombardment, and found that the bolometric $L_b$ agrees well
with equation (1) predicted by the PSG model. For completeness, in this
paper we include yet another pulsar PSR B1929$+$10, whose bolometric
thermal luminosity was recently determined by Misanovic, Pavlov \& Garmire
(2007). We show that this pulsar also satisfies equation~(1) by finding a
suitable feature in the modulation spectra data base of W06a,b (see
section 3.3 for some details). The number of pulsars satisfying and/or
being consistent with equation~(1) increased to seven. To the best of our
knowledge no single counter-example exits. It is worth emphasizing that
only pulsars for which both the bolometric luminosity $L_b$ of thermal
X-rays from hot polar cap and circulational periodicity $P_4$ of drifting
subpulses observed in radio band are known, can be used for this analysis.
In our sample of 8 available cases, 4 pulsars (B0656$+$14, B1055$-$52,
B0834$+$06 and B1929$+$10; see footnote 8 related to the latter case) and
3 others either show an evidence of hot spot thermal emission (B1133$+$16
and B0628$-$28) or at least such component cannot be excluded
(B0943$+$10). The last case (B0826$-$34) is uncertain as we only have an
upper limit for X-ray detection (consistent with PSG model).

\section{New X-ray data}
We have observed two radio pulsars B0834$+$06 and B0826$+$34 known for their prominent subpulse drift with the \xmm\
observatory (Jansen, Lumb, Altieri et al. 2001). We marked them in red color in Figure~1, to distinguish them from
previously analyzed four pulsars (marked in black) in Papers I and II. Yet another pulsar B1929$+$10 (marked in blue in
Figure~1) is discussed in Section~3.3, as its values of $L_b$ and $P_4$ have became recently available.

\subsection{PSR B0834$+$06}

The pulsar PSR B0834$+$06 was observed with \xmm\ on 2007 November 17 and
18 for a total of $\sim$71.7 ks. The EPIC-MOS (Turner, Abbey, Arnaud et
al. 2001) and EPIC-PN (Str{\"u}der, Briel, Dennerl et~al. 2001) cameras
were operated in imaging mode (see Table~\ref{tab-obs}). The observation
was scheduled at the end of the satellite revolution and the detector
background strongly increased when the satellite entered the radiation
belts. To maximize the signal to noise ratio we rejected the period of
high background which resulted in net exposure times around 50 ks
(Table~\ref{tab-obs}).

For the X-ray analysis we used the \xmm\ Science Analysis System (SAS) version 7.1.0 together with XSPEC version 11.3.2p for spectral modelling.
Standard SAS source detection based on a maximum likelihood technique was simultaneously applied to the X-ray images obtained from the three EPIC
instruments and five different energy bands (band B1 0.2$-$0.5 keV, B2 0.5$-$1.0 keV, B3 1.0$-$2.0 keV, B4 2.0$-$4.5 keV and B5 4.5$-$12.0 keV). A
weak source was found at the position of the pulsar at R.A. = 08 37 05.71 and Dec. = 06 10 15.8 (J2000.0) with a 1$\sigma$ statistical error of
1.7\arcsec. Nearly 150 X-ray sources were detected in the EPIC images and a comparison with catalogues from other wavelength bands yields many
correlations within $\sim$0.5\arcsec\ of the X-ray positions. This demonstrates that the systematic uncertainty in the astrometry is small compared
to the statistical error of the source position. The X-ray source position is within 1.3\arcsec\ of the radio position of PSR B0834$+$06, consistent
within the errors. The positional agreement and other properties of the X-ray source (see below) make a chance coincidence very unlikely.

The total EPIC count rate (summed for the three instruments in the 0.2$-$4.5 keV band) obtained from the source detection analysis is
(1.4$\pm$0.3)$\times$10$^{-3}$ \ct, insufficient for a detailed spectral analysis. To obtain constraints on the shape of the X-ray spectrum we
therefore use hardness ratios (X-ray colours) derived from the count rates in the standard energy bands and compare them with those expected from
various model spectra. Because the EPIC-PN detector is more sensitive, in particular at low energies where most of the counts are detected, we use
only count rates obtained from EPIC-PN. Hardness ratios are defined as HR1 = (R2-R1)/(R2+R1), HR2 = (R3-R2)/(R3+R2), HR3 = (R4-R3)/(R4+R3) and HR4 =
(R5-R4)/(R5+R4) with RN denoting the source count rate in band BN. To compare the measured hardness ratios with those inferred from model spectra, we
simulated expected EPIC-PN spectra (using XSPEC and the appropriate detector response files) and derived expected count rates and hardness ratios.

The distance to PSR B0834$+$06 estimated as 643 pc was derived from its dispersion measure of DM = 12.86 pc cm$^{-3}$ (from
the online ATNF pulsar catalog)\footnote{http://www.atnf.csiro.au/research/pulsar/psrcat/}. Assuming a 10\% ionization
degree of the interstellar matter along the line of sight to PSR B0834$+$06, this converts to a hydrogen column density of
\nh\ = 4.0\hcm{20}. Because of the low statistical quality of the X-ray data we are not able to derive tight constraints on
the absorbing column density. Therefore, we limit our investigated model parameter space to \nh\ values between 1.0\hcm{20}
(a lower limit which is reached within a distance of 200 pc; Posselt, Popov, Haberl, et al. 2008) and 8.0\hcm{20} (allowing
an uncertainty of a factor of 2 in the assumed ionization degree for the conversion from DM to \nh).

As model spectra we tested power-law (PL hereafter) and blackbody (BB hereafter) emission and a combination of the two. In
all model spectra absorption was included, assuming elemental abundances from Wilms, Allen \& McCray (2000). For the
absorbed power-law model we explored the parameter space for \nh\ between 1.0\hcm{20} and 8.0\hcm{20} with a step size of
1.0\hcm{20} and for the photon index $\gamma$ between 1 and 5 in steps of 0.2. Figure~2a shows the hardness ratios HR1
versus HR2 derived at the parameter grid points. The measured hardness ratios HR1 and HR2 are drawn with 1$\sigma$ (solid
lines) and 2$\sigma$ (dotted lines) error bars. The rectangular boxes around the error bars indicate the corresponding
confidence areas, although these are in reality limited by error ellipses which fit inside the boxes. As can be seen, the
power-law model spectra can not reproduce the measured hardness ratios within their 1$\sigma$ errors. Allowing 2$\sigma$
errors would require a relatively steep power-law with a photon index between 2 and 4 and preferentially high absorption.

The results for a BB model with temperatures varying between $kT_{\min} =
80$ eV and $kT_{\max} = 480$ eV in steps of 20 eV (\nh\ grid as above) are
plotted in Figure~2b. The measured hardness ratios are best reproduced by
the model with \nh\ = 4\hcm{20} and $kT = 170$ eV. The $1\sigma$
($2\sigma$) confidence range for the temperature is $kT = 170^{+65
(+120)}_{-55 (-80)}$ eV. We determine bolometric luminosity using the
model parameters at the grid points (normalizing the simulated spectra to
match the observed count rate in the 0.2$-$4.5 keV band) which yielded
\lbol\ = 8.6$^{+7.6 (+14.2)}_{-2.0 (-2.0)}$ \ergs{28}. It is remarkable
that the hydrogen column density derived from the ``best-fit'' BB model of
4.0\hcm{20} is fully consistent with the DM assuming 10\% ionization along
the line of sight to PSR B0834$+$06.

We also investigated a combination of BB and PL (with a photon index of 2.0 as typically seen in the X-ray spectra of
pulsars (e.g. Kargaltsev et al. 2006), both subject to the same absorbing column density. As first case the normalization of
the power-law component was set to have a flux (for the 0.2$-$10.0 keV band) in the PL component of 50\% of that in the BB
component, i.e. a flux ratio of \fbb:\fpl = 1:0.5. The hardness ratios are shown in Figure~2c. As expected, HR2, which is
most sensitive to the shape of the intrinsic spectral shape, increases with respect to the case of the pure BB due to the
contribution of the harder PL component. The $1\sigma$ ($2\sigma$) confidence ranges are $kT = 140^{+85 (+190)}_{-35 (-50)}$
eV and $\lbol\ = 9.9^{+4.3 (+8.6)}_{-4.4 (-4.4)}$ \ergs{28}. It should be noted here, that the luminosity of the BB
component increases, although a power-law component is added to the model spectrum. This is because the power-law rises
toward the low energies and a higher \nh\ values is required to compensate for that. A higher \nh\ in turn increases the
bolometric luminosity of the BB component in order to match the observed spectrum (hardness ratios and count rates) again.
These effects are also evident in the second case, where we used a flux ratio of \fbb:\fpl = 1:1 (Fig.~2d): HR2 increases
further and the upper limits for \lbol\ also rise somewhat ($kT = 140^{+80 (+210)}_{-40 (-55)}$ eV; \lbol\ = 9.9$^{+5.4
(+10.9)}_{-4.4 (-4.4)}$ \ergs{28}).

The above results are summarized in Figure~3 which presents $L_b$ versus
$kT$ obtained from the modelled hardness ratios in Figures~2b-2d, where
symbols (circle and square for $1\sigma$ and $2\sigma$ levels,
respectively), their colors (red, blue and green for BB, BB(2/3)+PL(1/3)
and BB+PL model, respectively) and related numbers, correspond to those
used in Figures~2b-2d. We can summarize that thermal radiation from the
hot polar cap of PSR B0834+06 is described by kT = (170$^{+180}_{-90})$ eV
(or surface temperature of the polar cap $T_s = (2.0^{+2.0}_{-0.9})\times
10^6$ K and $L_b$ =$(8.6^{+14.2}_{-4.4)}\times 10^{28}$~erg~s$^{-1}$,
where the $2\sigma$ errors are determined by both statistical and model
uncertainties.

\subsection{PSR B0826-34}
The pulsar PSR B0826-34 was observed with \xmm\ on 2006 November 13 and 14
with the EPIC-MOS and EPIC-PN cameras operated in imaging mode
(Table~\ref{tab-obs}). Also during this observation, strong background
flaring activity ocurred near the end of the observation. After background
screening a total exposure time of $\sim$ 38.8 ks was obtained.

We selected this source because it was one of the few pulsars with known $P_4$ value (Gupta, Gil, Kijak et al. 2004; G04
hereafter). When applying for the \xmm\  observing time we realized that PSR B0826-34 would be at most a very weak source
like PSR B0943+10 (or even weaker) detected by Z05. Indeed the spin-down value was quite low and even our equation~(1)
predicted the source luminosity twice lower than that of B0943+10. However, B0826-34 is closer to the Earth than B0943+10 by
the factor of 1.5. Despite relatively large DM=52.9 pc cm$^{-3}$ we optimistically assumed that the hydrogen column density
\nh\ will be similar to that of PSR B0943+10 (with DM=15.4 pc cm$^{-3}$). We speculated that the factor of 3.5 in DM values
would be compensated to some degree by the factor of 0.67 in a distance. We did not detect the pulsar, which probably means
that the actual value of \nh\ is much higher than assumed, due to some dense cloud of hydrogen along the line of sight to
B0826-34. Therefore, we determined the upper limit for thermal X-ray radiation from hot PC from this pulsar.

Because of the higher sensitivity of the EPIC-PN camera we used images from this instrument only. We created images in the energy bands $0.2-0.5$
keV, $0.5-1.0$ keV and $1.0-2.0$ keV and determined $2\sigma$ upper limit count rates for the expected source position for each energy band. The
total $0.2-2.0$ keV upper limit was obtained as 2.3 cts s$^{-1}$. Assuming the BB model with $kT = 267$ eV and the absorption column density of
3\hcm{20} ( 1.4\hcm{21}), this converts into an upper limit for the bolometric luminosity of \lbol\ = 1.0\ergs{29} (\lbol\ = 1.45\ergs{29}). The
latter value was conservatively used in Figure 1 (Table 1).

\section{Data analysis and model verification}

Table~1 and Figure~1 present the observational data of a number of quantities for seven pulsars, in which both $P_4$ and $L_b$ are known or at least
constrained. These data are confronted with the model curve representing equation~(1), which is marked by the solid line, accompanied by broken lines
describing theoretical errors due to uncertainty with determination of the neutron star moment of inertia (see Appendix~A). Two pulsars: B0943+10 and
B1133+16 have already been discussed in Paper I, and two others B0656+14 and B0629$-$28 in Paper II. As argued in Papers I and II these pulsars
strongly support our theory (they are presented as black dots in our Figure 1). In the following we study and discuss the results from the remaining
three pulsars: B0834+06, B0826$-$34 (red dots) and B1929+10 (blue dot).

\subsection{PSR B0834$+$06}
As already mentioned, the circulational (tertiary) period $P_4$ is known for a handful of pulsars, and B0834+06 is one of
them. The first measurement of tertiary periodicity for this pulsar was made by Asgekar \& Deshpande (2005; AD05
henceforth), who argued that $P_4/P=15 \pm 0.8$ and the number of circulating sub-beams (sparks) $N=P_4/P_3=8$, implying the
aliased subpulse drifting with primary period $P_3/P=1.88 \pm 0.01$. They found a strong low frequency feature in the
intensity fluctuation spectrum at $0.07~c/P$ in one sequence of 64 single pulses, supported by side tones flanking the
primary feature of $0.46~c/P$ by $\pm 0.066~c/P$. These results seemed quite robust, although a small derived number of
sparks (8) as compared with other cases was a bit worrying. We used $P_4/P=15$ in the scientific justification for \xmm\
proposal, predicting from equation~(1) a quite luminous hot PC in PSR B0834+06, emitting with \lbol\ = 36\ergs{28}. The
model simulations indicated the count rate of about 0.018 cts~s$^{-1}$, implying a very promising case. Slightly before the
scheduled XMM observing session a new estimate was obtained by Rankin \& Wright (2007; RW07 henceforth), who argued, using
their new Arecibo data and new technique involving a distribution of null pulses, that $P_4/P \sim 30.25$. They argued that
the number of sparks and/or subbeams involved in the non-aliased subpulse drift with the true primary period $P_3/P=2.16 \pm
0.01$ is 14 and thus $P_4/P=30.24 \pm 0.15$ (Table 1). According to equation~(1) this would imply the luminosity 4.16 times
lower than \lbol\ = 36\ergs{28} given in our proposal, that is \lbol\ = 8.85 \ergs{28} or $L_b/\dot E=0.67\times 10^{-3}$.
Amazingly, this is almost exactly the central value of our best fit for hot BB component in PSR B0834$+$06 (see Table 1 and
Figs.~1 and 3). Thus our measurements interpreted within the PSG model (eq.~[1]) strongly support the value of $P_4/P=30.25
\pm 0.25$ obtained by RW07, while $P_4/P=15 \pm 0.8$ obtained by AD05 is highly unlikely.

\subsection{PSR B0826$-$34}
The carousel rotation time in this pulsar was obtained by means of
computer simulations compared with real single pulse data by Gupta, Gil,
Kijak et al. 1984. According to equation (1) its valule $P_4=(14 \pm 1)P$
implies the efficiency $L_b/\dot E=3.2\times 10^{-3}$. These values are
marked by the red horizontal error bar labelled by B0826-34. The upper
limit 22 $\times 10^{-3}$ is marked as the short arrow above. This pulsar
would have to be much more efficient in converting the spin-down power
into X-rays to be detected in a 50 ksec \xmm\  exposure, or at least a six
times longer exposure time would be required.

\subsection{PSR B1929$+$10}
Recently Misanovic, Pavlov and Garmire (2007; M07 hereafter) argued that
X-rays from PSR B1929$+$10 include both magnetospheric and thermal
components. The BB fit to the latter gives a temperature k$T$=0.3 keV and
a projected surface area $A_p \sim 3.4 \times 10^3$~m$^2$ or radius $r_b$
of about 33 meters (much smaller than the canonical $A_{pc} = 2 \times
10^5$~m$^2$ or $r_b \sim 300 ~meters$). This corresponds to the bolometric
luminosity $L_b \sim (1-2) \times 10^{30}$ ~ergs s$^{-1}$ emitted from hot
($T=3.5 \times 10^6$ ~K) polar cap with a radius of about 33 meters.
\footnote{Recently, Hui \& Becker (2008; henceforth HB07) analyzed the
same XMM-Newton data of B1929$+$19 (using different way of data binning
resulting in better photon statistics per spectral bin) and argued that
the hot BB component is statistically unjustified. However, if they
allowed the BB radius and temperature of the hot spot as the free
parameters, then the best fit resulted in very small hot spot area with a
radius $r_b=25.81^{+18.81}_{-25.81}$ meters, perhaps even smaller that the
one obtained by M07. In opinion of HB08 this is unacceptable small as
compared with the canonical PC radius. However, within our model this is a
result of relatively low dipolar surface magnetic field $B_d=5 \times
10^{11}$ Gauss. The actual non-dipolar magnetic field must be much higher
(about 400 times) to provide enough binding energy (ML07) for creation of
the PSG in this pulsar, which results in the hot spot radius
$r_b=300/20=15$ meters (see section 5 for more details).} We used the
central value of B1929$+$10 $L_b=1.17^{+0.13}_{-0.4}$ ~ergs s$^{-1}$ with
2$\sigma$ errors from M07 (see the top panel in their Figure~11).

For each new pulsar with a known value of thermal bolometric luminosity $L_b$ we search the available data bases for a possible value of $P_4$. In
case of PSR B1929+10 we found in W06a (their Figure~A13) a clear but weak low frequency spectral feature at about 0.02 $c/P$. This translates into a
long periodicity $P_4/P = 50^{+15}_{-5}$, with errors estimated from half-width of the low frequency feature. Going back to Figure~1 we see that the
data point (marked in blue color) for B1929+10 (Table 1) fits the theoretical curve very well. This is an important point, as it extends the
parameter space to the low efficiency/(long period) region in our Figure 1. The range of parameters for our 7 cases under examination increased to
factors of ~67 and ~7 for the efficiency $L_b/\dot E$ and the tertiary period $P_4/P$, respectively.

\subsection{PSR B1055$-$52} This is a bright radio pulsar showing complex
patterns of single pulse intensity modulations. The drifting subpulses are
not apparent but this can be the result of a central cut of the
line-of-sight (LOS) throughout the emission beam. Indeed, this pulsar has
a strong interpulse (IP) separated from the main pulse (MP) by about 145
degrees of longitude (measured between centroids) and both these
components have complex profiles, consistent with central LOS traverse.
B90 analyzed the fluctuation spectrum and found in part of the profile a
small and broad feature at frequency 0.045 cycles/$P$ with $Q \sim 1.5$.
This frequency and low Q can be interpreted as the carousel periodicity
$P_4/P=22^{+11}_{-5}$. Recently, Mitra (2008) confirmed this feature at
parts of both MP and IP in his data taken at GMRT (privat information).

The pulsar PSR B1055$-$52 is a luminous source of X-ray emission. DL05
identified three spectral components in this radiation: Power law
magnetospheric emission, cool BB emission from the entire surface of the
cooling neutron star, and hot BB emission from a small hot spot. This
latter component is of special interest for us and its parameters along
with references are listed in Table 1. If the "carousel" hypothesis
discussed above is correct, then we expect a correlation between the
carousel period $P_4$ and the bolometric luminosity $L_b$ from the hot
spot, according to our equation~(1). As shown in Table 1 and Fig. 1, the
bolometric luminosity $L_b=(1.6^{+0.9}_{-0.4})10^{31}$ erg/s, where the
errors were estimated from fitting the EPIC-pn spectrum, extracted from
the XMM-Newton archival data, with the same model as used by DL05.
Although the central values result in the data point lying slightly below
the theoretical curve, PSR B1055$-$52 is certainly consistent with
equation (1). Indeed, one can see that values slightly higher than the
central one, e.g., $P_4/P \sim 28$ and $L_b/\dot E \sim 0.8 \times
10^{-3}$ would result in a very good fit to the equation (1).

\section{Conclusions and Discussion}

Within the partially screened gap (PSG) model of the inner acceleration
region in pulsars developed by G03, we derived in Paper I a simple and
clean relationship (eq.~[1]) between the thermal X-ray bolometric
luminosity $L_b$ from hot PC heated by sparks and the circulation time
$P_4$ of the spark-associated drift detected as the subpulse drift in
pulsar radio emission. This relationship expresses the well justified
assumption (Appendix A) that both the drifting rate and the polar cap
heating rate are determined by the same value of electric field within the
inner acceleration region. Indeed, the drifting rate described by
measurable $P_4$ is determined by the tangent (with respect to surface
magnetic field) component of the electric field, while the heating rate
described by measurable $L_b$ is determined by its component parallel to
the surface magnetic field in the (partially screened) gap. In Paper II we
showed that PSRs B0943$+$10, B1133+16, B0628$-$20 and B0654+14, which were
the only pulsars with both $L_b$ and $P_4$ known at that time, satisfied
equation~(1) quite well (see also Fig.~1 and Table~1). This suggested that
the PSG model may indeed be a reasonable description of the inner
accelerator region near the polar cap. In this paper we support this view
by demonstrating that another two pulsars (B0834+06, B1929+10 and
B1055$-$52) also satisfy the equation~(1). Yet another pulsar B0826-34, in
which only the upper limit for $L_b$ was obtained, demonstrated a
consistency with equation~(1) as well.

Only for a handful of pulsars the circulation (carousel) time was measured or constrained so far. Measurement of $P_4$ by means of modulation
spectral analysis requires a strong unevenness in the circulating system, maybe a distinguished group of adjacent sparks or even just a single spark
(see also the scenario discussed by Gil \& Sendyk, 2003; GS03 hereafter). Moreover, this feature should persist considerably longer than the
circulation time. Such favorable conditions do not occur frequently in pulsars and therefore direct or indirect measurements of $P_4$ are very rare.
In principle, in a clean case one should be able to detect the primary feature $P_3$, reflecting the phase modulation of regularly drifting
subpulses, flanked by two symmetrical features corresponding to slower amplitude modulation associated with carousel circulation as well as direct
low frequency feature $1/P_4$ (like in the case of PSR B0943+10; DR01, AD01 and GS03). However, results of Paper II clearly showed that $P_4$ can be
found also in pulsars without regularly drifting subpulses (and/or in erratic drifting modes). This strongly suggested that no matter the degree of
the organization of spark plasma filaments at the polar cap, the slow circumferential plasma drift was always performed at about the same rate in a
given pulsar. The problem was how to reveal this motion. Two new methods were discussed or at least mentioned in Paper I. The 2-D phase resolved
modulation spectral analysis developed by ES02 and ES03 and implemented by W06a, b was the first one. The second method based on examination of the
distribution of nulls in the long sequence of single pulses was recently developed by HR07 and Rankin and Wright (2007; RW07 henceforth). In view of
the main results obtained in this paper the latter method deserves some more detailed discussion here.

As discussed in section 3.1 there is a controversy about the actual value
of $P_4$ in PSR 0834+06. AD05 reported that the alias-corrected
$P_3/P=1.88 \pm 0.01$ and $P_4/P=15 \pm 0.8$, implying the number of
sparks $N=P_4/P_3=8$. These authors found just one sequence of 64 pulses
in which the fluctuation spectrum analysis revealed the low frequency
feature at about 1/15=0.067. On the other hand RW07 found the non-aliased
primary drift periodicity $P_3/P=2.16 \pm 0.011$ and the number of sparks
$N=15$, implying the tertiary long periodicity $P_4/P=30.24 \pm 0.15$.
This longer cycle with $P_4 \sim 30 P$ was supported by our measurements
of $L_b$ and PSG model expressed by equation~(1). RW07 examined an
interaction between nulls and emission in PSR B0834+06. They found that
null pulses are not randomly distributed and that the most likely
periodicity in their appearance is about $30 P$. Following the previous
discovery of HR07 that null pulses and drifting subpulses in PSR B1133+16
are associated with the same long periodicity (about $33 P$) RW07
convincingly argued that short pseudo-nulls (one pulsar period or less)
are just a result of irregular distribution of subpulse subbeams/sparks
that persist on time scales of at least hundreds of pulsar periods. The
short-time pseudo-nulls appear when the line-of-sight cuts through the
low-level emission region in the radio beam. Our results on both B1133+16
and B0834+16 strongly support this picture. The interesting question is
then why AD05 obtained such a strong feature at $15 ~P$ for a sequence of
64 single pulses from B0834+06. RW07 admitted that they also found in
their data some sequences showing $15P$ periodicity, which seemed to be a
sub-harmonic of $30P$ cycle. We noticed yet another problem with the
result of AD05. In our opinion, these authors have used incorrectly their
equations~(2) and (3). In fact, as $\Delta \phi$ they used the
longitudinal distance between the profile components and in consequence,
the azimuthal magnetic angle between the neighboring subbeams was $\Delta
\theta=50$ degs.  This ignored the subpulses appearing in the saddle of
the profile. We believe that they should use $\Delta \theta \sim 25$ degs,
and as a result, the number of sparks would be $N=360/25=14$ instead 8.
This is consisted with $P_4=N P_3=14\cdot 2.16=30.24 P$ obtained by RW07
and supported by our results presented in this paper. In summary, we
strongly believe that the actual value of $P_4$ in PSR B0834+06 is close
to 30 pulsar periods and that $15 P$ corresponds to a first harmonic of
the basic cycle. Some evidence of low frequency spectral features at both
$0.033~c/P$ and $0.066~c/P$ can be seen in Figure~A19 of W06. Moreover, it
seems that 14 sparks inferred by RW07 are more likely than 8 sparks
inferred by AD05.

The essence of the PSG pulsar model is the presence of a strong,
nondipolar surface magnetic field $B_s$, although it does not appear
explicitly in equation (1); see Appendix A for details. The strong value
of $B_s$ is necessary for providing enough binding (cohesive energy) to
prevent the free flow of iron ions from the surface (Medin \& Lai 2007;
ML07 hereafter), while the small radius of curvature is needed to develop
cascading pair production (e.g. Gil \& Melikidze 2002). The latter
phenomenon is essential for both shorting out the gap potential drop and
providing a dense electron-positron plasma in the radio emission region
(eg. Melikidze \& Gil, 2000 and Gil, Lyubarski \& Melikidze, 2004). When
the calculations of ML07 are adapted to the PSG model, then one can derive
the dependence of the surface magnetic field on the surface temperature
$B_s=B_s(T_s=T_i)$; (we will give detailed description of this topic in a
separate paper, but see Appendix A for some details). For the condensed Fe
surface this relationship is represented by the solid red line in Figure 7
of ML07. We can apply this apparatus to our case of PSR B0834+06, with
$L_b=(6.8^{+1.1}_{-1.3})\times 10^{28}$ erg~$s^{-1}$,
$T_s=(2.0^{+2.0}_{-0.9})\times 10^{6}$ K, and the associated effective
surface area of the hot spot $A_{p}=940~$ m$^2$. On the other hand, one
can read off from Figure 7 in ML07 the range of values $B_s \sim
(1^{+1.3}_{-0.6})\times 10^{14}$ G corresponding to
$T_s=(2.0^{+2.0}_{-0.9})\times 10^6$~K. Since the dipolar surface magnetic
field and polar cap area are $B_d=3 \times 10^{12}$ G and $A_{pc}=4.85
\times 10^{4}$ m$^2$, respectively, we can find the effective surface area
$A_p=A_{pc}B_s/B_d=(1.5^{+1.4}_{-0.9})\times 10^{3}$ m$^2$. This is
consistent with our estimate, in which $A_p$ is about 50 smaller than
$A_{pc}$. Theoretically, this results naturally from the flux conservation
of the open magnetic field lines. As pointed out in Paper II (see also
references therein), the small size of the hot spot relative to the
canonical polar cap area is a typical property of hot BB thermal radiation
detected in a number of pulsars. An extreme case was published just
recently by Pavlov, Kargaltsev, Wong et al. (2008; P08 hereafter), who
reported on the Chandra detection of a very old (170 Myr) and close to the
Earth (0.13 kpc and 0.184$^{+0.01}_{-0.017}$ ~kpc, according to ATNF
(Manchester, Hobbs, Teoh et al. 2005) and NE2001 (Cordes \& Lazio 2003)
database, respectively) radio pulsar PSR J0108-1431, with a very weak
dipolar surface magnetic field $B_d=2.52 \times 10^{11}$ ~G and a low
spindown $\dot {E}=5.8 \times 10^{30}$ erg~$s^{-1}$. During 30 ks exposure
they detected 53 counts and found that the spectrum can be described by PL
model or BB model equally well. For the latter model they obtained the
bolometric luminosity $L_b=1.3 \times 10^{28} d^{2}_{130}$~erg~$s^{-1}$,
$T_s=3.2 \times 10^{6}$~K and $A_p=50~d^{2}_{130}$~m$^2$, which translates
into the hot spot radius as small as 4 meters. This is the smallest hot
polar cap ever observed$^8$, with the ratio $b=A_{pc}/A_p=1.77 {\times
10^3}/d^{2}_{130}$, equal to 1770 or 923 (highest ever obtained) for
distances 0.13 and 0.18 kpc, respectively . Accordingly, the actual
surface magnetic field $B_s=bB_d$ (see Gil \& Sendyk 2000 and ML07) is
equal to 4.5 or 2.3 $\times 10^{14}$~G for a distance of 0.13 or 0.18 kpc,
respectively. Interestingly, the latter value agrees almost exactly with
ML07 (red solid line in their Fig. 7), while the former one implies too
high a surface temperature exceeding 5 MK. Thus, the extremely small hot
polar cap with $T_s=$3.2 MK results from the fact that the actual surface
magnetic field must be about 1000 times stronger than the dipolar
component, in order to provide enough cohesive energy to develop PSG in
this pulsar. We can therefore say that the case of PSR J0108-1331 supports
strongly the PSG pulsar model, the ML07 cohesive energy calculations for
the condensed Fe polar cap surface and NE2001 distance to this pulsar
(about 0.184 kpc). If one adopts 0.184 kpc as the proper distance to PSR
J0108-1331, then the bolometric BB luminosity is $L_b \sim 2.5 \times
10^{28}$ erg~$s^{-1}$ and the efficiency $L_b/\dot {E} \sim 4.3 \times
10^{-3}$. With this value the equation (1) predicts the tertiary
periodicity $P_4/P \sim$~12. However, the confirmation of this by means of
single pulse radio observations of PSR J0108-1431 seems hopeless with
present day possibilities, as the pulsar is also extremely weak in radio
band (Tauris, Nicastro, Johnston et al. 1994).

Thus, our PSG model seems to account for the physical phenomena at and
above the actual pulsar polar cap quite well. Other available inner
acceleration models do not match the observations well. The pure vacuum
gap model (Ruderman \& Sutherland 1975) has $\eta=1$. Although it also
satisfies Eq.(1), it predicts a very high polar cap heating rate,
typically $L_b \sim 10^{-1}-10^{-2} \dot E$ (Zhang et al. 2000), and
therefore a very small $P_4$. The predicted high $L_b$ has been ruled out
by the X-ray observations of many old pulsars (ZSP05, TO05, K06 and this
paper), and the predicted low $P_4$ is also inconsistent with the radio
observations. On the other hand, as discussed in \S1 the steady-state SCLF
model does not predict the existence of the ``sparks'' whose drifts around
the polar cap region provide the most natural interpretation of the
observed drifting sub-pulse patterns. A modified unsteady SCLF model
(which has not been discussed in the literature) may be able to introduce
a sparking-like behavior. Based on the similar logic (i.e. the potential
drop along the magnetic field line in the gap is equal to the horizontal
potential drop across the spark, see Appendix), a similar equation as
Eq.(1) can be derived for the SCLF model. However, since this model
introduces a very small effective $\eta$ value ($\eta \sim (2\pi
R_*/cP)^{1/2} << 1$, Harding \& Muslimov 2001), the predicted polar cap
heating rate is too low to interpret the observations, typically $L_b \sim
10^{-4}-10^{-5} \dot E$ (Harding \& Muslimov 2002). Also the corresponding
drifting velocity is too small so that the predicted $P_4$ is too long as
compared with the radio data. The PSG model predicts an intermediate
particle inflow rate, and gives the clean prediction (Eq.[1]) which allows
$L_b$ to be a moderate value. This is strongly supported by the data.

In order to solve the binding energy problem in the canonical dipolar
magnetic field at the neutron star surface, it has been conjectured that
drifting subpulse pulsars are bare strange stars (Xu et al. 1999). The
simplest model does not allow a hot polar cap because of the high thermal
conductivity of the bare strange star surface layer, which is ruled out by
the data. Yue et al. (2006) argued that PSR B0943$+$10 may be a low mass
quark star ($\sim 0.02 {\rm M}_\odot$). However, pulsar drifting seems to
be the most common behavior of radio pulsars (W06a,b), some of which have
well measured mass around $1.4 {\rm M}_\odot$ (Thorsett \& Chakrabarty
1999). We regard that the quark star scenario is no longer attractive in
view of the latest observations. The cohesive energy calculations of Fe
ion chains in ultra-strong magnetic field by ML07 seem to be strongly
supported by the X-ray observations discussed in this paper.

Finally, we would like to address a hypotheses put forward by Becker,
Kramer \& Jessner et al. (2006) that in old pulsars ($>10^6$ yrs) the
magnetospheric emission dominates over thermal emission, including both
cooling radiation and hot polar cap emission component. These authors
suggested that the latter radiation component decreases along with the
former one, and if so, the hot polar caps in cooling neutron stars could
be formed by anisotropic heat flow due to the presence of the magnetic
field rather than by particle bombardment. While in young NSs with core
temperature $\simeq 10^8$~K the strong crustal magnetic fields may channel
the heat toward the polar cap resulting in $T_s$ of a few MK
(Perez-Azorin, Miralles \& Pons 2006; Geppert, K{\"u}ker \& Page 2006), in
pulsars older than $10^6$ years this mechanism is much less efficient and
the only viable process that can produce such hot and small polar caps is
the back-flow particle bombardment. Almost all pulsars presented and
examined in this paper are older than 1 Myr (an exception is 110 kyr PSR
B0656+14). For instance, PSR B0834+06 is 3 Myr old and its X-ray emission
is dominated by hot BB component (an obvious counter-example arguing
against Becker's claim). In PSR B1929+10 (3.1 Myr old) the luminosity of
hot BB component is at least comparable with the magnetospheric X-ray
radiation (M07). The very old (170 Myr) rotation powered non-recycled
pulsar J0108-1431 clearly shows BB radiation from the hot polar cap (P08),
probably accompanied by the magnetospheric emission, but no evidence of
cooling radiation from the whole surface, as expected for such an old
pulsar.

In summary, both the polar cap full cascade (Zhang \& Harding 2000) and
the downward outer gap cascade (Cheng, Gil \& Zhang 1998) that have been
proposed to interpret non-thermal X-ray emission from spindown-powered
pulsars are expected to be less significant in pulsars from our sample
with respect to the young pulsars. The predicted values of X-ray
luminosity in these models are typically lower than that of the polar cap
heating in the PSG model (Eq.[1]). In view that other available models of
the pulsar inner accelerator (pure vacuum gap model and
space-charge-limited flow model) either overpredict or underpredict the
polar cap heating level, we conclude that the pulsar inner accelerator is
likely partially screened due to a self-regulated sub-Goldreich-Julian
flow. Also, the pure vacuum gap model predicts too fast a drifting and the
space-charge-limited flow model has no natural explanation for the
subpulse drift phenomenon at all. We thus strongly believe that thermal
radiation associated with a polar cap heating due to partially screened
inner accelerator (PSG) is a common component of pulsar X-ray emission
regardless of its age, and this component plays especially significant
role in the spectra of old pulsars.

\acknowledgements
Our results are partly based on observations with
XMM-Newton, an ESA Science Mission with instruments and contributions
directly funded by ESA Member states and the USA (NASA). We acknowledge
the support of the NASA NNX07AF07G and NNX08AC67G grants. JG was partially
supported by the Polish State Committee for Scientific Research grant N
N203 2738 33 and GM was partially supported by the Polish State Committee
for Scientific Research grant N N203 1262 33, as well as by the Georgian
NSF ST06/4-096 and INTAS 06-1000017-9258 grants. The \xmm\ project is
supported by the Bundesministerium f\"ur Wirtschaft und
Technologie/Deutsches Zentrum f\"ur Luft- und Raumfahrt (BMWI/DLR, FKZ 50
OX 0001) and the Max-Planck Society. We thank Dr. Dipanjan Mitra for
stimulating discussions, critical reading of the manuscript and helpful
comments.

\appendix

\section{Inner acceleration region in pulsars}

The charge depleted inner acceleration region above the polar cap results from the deviation of a local charge density $\rho$ from the co-rotational
charge density (Goldreich \& Julian 1969) $\rho_{\rm GJ}=-{\mathbf\Omega}\cdot{\bf B}_s/{2\pi c}\approx{B_s}/{cP}$. For isolated neutron stars one
might expect the surface to consist mainly of the iron formed at the neutron star's birth (e.g. Lai 2001). Therefore, the charge depletion above the
polar cap can result from binding of the positive $^{56}_{26}$Fe ions (at least partially) in the neutron star surface. If this is really possible
(see Medin \& Lai 2006, 2007 and Paper II for details), then the positive charges cannot be supplied at the rate that would compensate the inertial
outflow through the light cylinder. As a result, a significant part of the unipolar potential drop develops above the polar cap, which can accelerate
positrons to relativistic energies and power the pulsar radiation mechanism, while the electrons would bombard the polar cap surface, causing a
thermal ejection of ions, which are otherwise more likely bound in the surface in the absence of additional heating. This thermal ejection would
cause partial screening of the acceleration potential drop $\Delta V$ corresponding to a shielding factor $\eta=1-\rho_{i}/\rho_{\rm GJ}$ (see GMG03
for details), where $\rho_{i}$ is the charge density of the ejected ions, $\Delta V=\eta({2\pi}/{cP})B_s h^2$ is the potential drop and $h$ is the
height of the acceleration region.  The gap potential drop is completely screened when the total charge density $\rho=\rho_i+ \rho_+$ reaches the
co-rotational value $\rho_{GJ}$. In terms of binding of $^{56}_{26}$Fe ions, the screening factor $\eta=1-exp(C_i-\varepsilon_c/kT_s)$,
$\varepsilon_c$ is the cohesive energy of the condensed iron surface, $T_s$ is the actual surface temperature, $T_i=\varepsilon_c/kC_i$ is the
critical temperature above which the iron ions are ejected with the maximum co-rotation limited rate, and $C_i=30 \pm 3$ (Medin \& Lai 2007).

Because of the exponential sensitivity of the accelerating potential drop
to the surface temperature, the actual potential drop should be
thermostatically regulated. In fact, when the potential drop is large
enough to ignite the cascading pair production, the back-flowing
relativistic charges will bombard the polar cap surface and heat it at a
predictable rate. This heating will induce thermionic emission from the
surface, which will, in turn, decrease the potential drop that caused the
thermionic emission in the first place. As a result of these two
oppositely directed tendencies, the quasi-equilibrium state should be
established, in which heating due to electron bombardment is balanced by
cooling due to thermal radiation. This should occur at a temperature
slightly lower than the critical temperature above which the polar cap
surface delivers thermionic flow at the corotational charge density level.
This is an essence of the PSG model. For practical reasons it is assumed
that $T_s=T_i$, while in reality $T_s$ is few thousands K lower than
$T_i$, with the latter being strongly dependent on the surface magnetic
field $B_s$. This is illustrated by Figure~7 in Medin \& Lai (2007), which
was prepared for the pure VG model. The PSG model is realized along the
red (for Fe) line in this figure, which shows that for a few MK surface
temperatures, as suggested by X-ray observations of pulsar hot spots (see
Paper II and references therein) the surface magnetic field must be close
to $10^{14}$ G in all pulsars. For most pulsars this is a much stronger
field than that inferred from pulsar spindown due to the magnetic dipole
radiation. Therefore, the surface magnetic field in neutron stars must be
dominated by crust anchored non-dipolar magnetic anomalies. Such strong
and curved surface magnetic field is also necessary for development of the
cascading pair production via curvature radiation (e.g. RS75, Gil \&
Melikidze 2002).

Several models proposed for generating pulsar radio emission based on the
concept of vacuum gaps need radius of curvature of surface magnetic field
much smaller than the stellar radius (see for e.g. Gil, Melikidze , Mitra
2002). A possibility of generating such fields would be from currents in
the neutron stars crust (e.g. Urpin, Levshakov \& Iakovlev 1986, Geppert,
Rheinhardt \& Gil 2003). Mitra, Konar, \& Bhattacharya (1999) examined the
evolution of multipole components generated by currents in the outer
crust. They found that mostly low-order multipoles contribute to the
required small radii of curvature and that the structure of the surface
magnetic field is not expected to change significantly during the radio
pulsar lifetime.

The spark plasma inside PSG must slowly drift with respect to the polar
cap surface due to non-corotational charge density. This drift will
manifest itself by the observed subpulse drifting, provided the spark
arrangement is quasi-stable over time scales of hundreds of pulses or so.
The deviation of the charge density from the co-rotational value generates
an electric field $\bf\Delta{\mathbf E}=\bf\Delta{\mathbf
E_\parallel+\bf\Delta{\mathbf E_\perp}}$ just above the polar cap surface.
The parallel component causes acceleration of charged particles, while the
perpendicular component participates in the subpulse drift. The tangent
electric field at the polar cap boundary, $\Delta E_\parallel=0.5{\Delta
V}/{h}=\eta({\pi}/{cP})B_sh$ (see Appendix~A in GMG03 for details). Due to
the $\bf\Delta{\mathbf E}\times{\mathbf B_s}$ drift the discharge plasma
performs a slow circumferential motion around the magnetic axis (see the
next paragraph below) with velocity $v_d=c\Delta E_\perp/B_s=\eta\pi h/P$.
The time interval to make one full revolution around the polar cap
boundary is $P_4\approx 2\pi r_p/v_d$. One then has
\be
\frac{P_4}{P}=2\frac{r_p}{\eta h\alpha} ~,
\label{P3P}
\ee
where the coefficient $\alpha=\Delta E_\perp/\Delta E_\parallel$ should be close to unity. If the plasma above the polar cap is fragmented into
filaments (sparks), which determine the intensity structure of the instantaneous pulsar radio beam, then in principle, the circulational periodicity
$P_4$ can be measured/estimated from the pattern of the observed drifting subpulses (Deshpande \& Rankin 1999, Gil \& Sendyk 2003). In practice,
$P_4$ is measured from the low frequency features in the modulation spectra obtained from good quality single pulse data of pulsars with drifting
subpulses. According to RS75, $P_4=NP_3$, where $N$ is the number of sparks contributing to the drifting subpulse pattern observed in a given pulsar
and $P_3$ is the primary drift periodicity (distance between the observed non-aliased subpulse drift bands).

The circumferential motion around the magnetic axis like in RS75 holds
only when the magnetic and the spin axes are almost parallel (almost
aligned rotator, in which the line-of-sight trajectory is almost the
circumferential tracks of sparks moving around the magnetic axis). Many
pulsars with drifting subpulses have indeed a very broad profile
characteristic of the almost aligned rotators: e.g. B0826-34, B0818-41.
Others, which are not a broad profile pulsars and show regular drifting
must have very high impact angle, i.e. grazing the emission beam. In such
cases one cannot exclude the almost aligned geometry. In more general
(inclined) case, the spark trajectory does not have to be closed on the
polar cap, as sparks should rather follow the trajectory of the
line-of-sight projected onto the polar cap, being slightly late behind the
star's rotation. However, observations of drifting subpulses in some
pulsars do not support such a scenario, being consistent with the
circumferential motion of the spark-associated sub-beams of subpulse
radiation, even if pulsar is not an aligned rotator. Indeed, an orderly
drifting subpulses always demonstrate a systematic intensity modulation,
either increasing or decreasing towards the pulse profile midpoint. Also,
in pulsars with more central cut of the line-of-sight trajectory the
subpulse drift is less apparent (or none) but a characteristic
phase-stationary modulation of subpulse intensity modulation persist.
These properties strongly suggest that sparks move on closed trajectories
on the polar cap, although they do not have to be circular, like in
axially symmetric RS75 model, to the extent that in some of the detection
of circumferential motion with specified value of $P_4$ periodicity is
possible. A good example of such pulsar with central light-of-sight cut is
B0834+06 discussed in this paper. There must be then some agency that
makes sparks moving across the the line-of-sight projection on closed
trajectories around the local magnetic pole instead around the rotational
pole, irrespective of the inclination and impact angles.

The quasi-equilibrium condition is $Q_{cool}=Q_{heat}$, where $Q_{cool}=\sigma T_s^4$ is the cooling power surface density
by thermal radiation from the polar cap surface and $Q_{heat}=\gamma m_ec^3n$ is the heating power surface density due to
back-flow bombardment, $\gamma=e\Delta V/m_ec^2$ is the Lorentz factor, $n=n_{GJ}-n_{i}=\eta n_{GJ}$ is the number density
of the back-flowing particles that deposit their kinetic energy at the polar cap surface, $\eta$ is the shielding factor,
$n_{i}$ is the charge number density of the thermionic ions and $n_{GJ}=\rho_{GJ}/e=1.4\times
10^{11}b\dot{P}_{-15}^{0.5}P^{-0.5}{\rm cm}^{-3}$ is the corotational charge number density and $\dot{P}_{-15}$ is the time
derivative of the period in $10^{-15}$. It is straightforward to obtain an expression for the quasi-equilibrium surface
temperature in the form $T_s=(2\times 10^6{\rm K})(\dot{P}_{-15}/{P})^{1/4}\eta^{1/2}b^{1/2}h_3^{1/2}$ (Paper II), where
$h_3=h/10^3 ~cm$, the parameter $b=B_s/B_d=A_{pc}/A_{p}$ (Gil \& Sendyk 2000, Medin \& Lai 2007) describes the domination of
the local actual surface magnetic field over the canonical dipolar component at the polar cap, and $\dot{P}_{-15}$ is the
normalized period derivative. Here $A_{pc}=\pi r^2_{pc}$ and $A_{p}=\pi r^2_p$ is the canonical (RS75) and actual emitting
surface area, with $r_{pc}$ and $r_p$ being the canonical (RS75) and the actual polar cap radius, respectively. Since the
typical polar cap temperature is $T_s \sim 10^6$ K (Paper II), the actual value of $b$ must be much larger than unity, as
expected for the highly non-dipolar surface magnetic fields.

Using equation~(\ref{P3P}) one can derive the formula for thermal X-ray
luminosity as
\be
L_b=2.5\times 10^{31}\alpha^{-2}\left(\frac{\dot{P}_{-15}}{P^3}\right)\left(\frac{P_4}{P}\right)^{-2}
\label{lx},
\ee
or in the simpler form representing the radiation efficiency with respect
to the spin-down power $\dot{E}=I\Omega\dot{\Omega}=3.95 I_{45}\times
10^{31}\dot{P}_{-15}/P^3$~erg/s, where $I=I_{45}10^{45}$g\ cm$^2$ is the
neutron star moment of inertia and $I_{45}=1^{+1.25}_{-0.22}$ (see Papers
I and II for details)

\be \frac{L_b}{\dot{E}}=0.63 \left(\frac{\alpha^{-2}}{I_{45}}\right)
\left(\frac{P_4}{P}\right)^{-2}
\label{Lx}.
\ee
This equation is very useful for a direct comparison with the
observations, since it contains only the observed quantities (although it
is subject to small uncertainty factors related to the unknown moment of
inertia $I_{45}$ and the coefficient $\alpha$. It does not depend on any
details of the sparking gap model like non-dipolar surface magnetic field
$b=B_s/B_d$, the height $h$ of the acceleration region and the shielding
factor $\eta$, since they cancel in the derivation procedure, as they
suppose to do so. Indeed, this equation reflects the fact that both the
subpulse drifting rate (due to $\bf\Delta{\mathbf E}\times{\mathbf B_s}$
plasma drift) and the polar cap heating rate (due to back-flow
bombardment) are determined by the same physical quantity, which is the
potential drop across the inner acceleration region just above the polar
cap. No other agency should be involved. In practical application of
equation (A3) we will set $I_{45}=1$ and $\alpha=1$. The former is
commonly used and the latter means that the values of the accelerating
$E_\parallel$ and perpendicular $E_\perp$ components of electric field in
the PSG are almost the same. It is quite a reasonable assumption, all the
more that it seems to be supported observationally (Fig. 1).

{}

\clearpage

\begin{deluxetable}{llllllllll}
\tabletypesize{\scriptsize}
\tablecaption{Thermal X-ray radiation for hot polar cap in pulsars with drifting subpulses. \label{tbl-1}}

\tablehead{
\colhead{PSR} & \colhead{P (s)} & \colhead{$\dot{P}_{-15}$} & \colhead{$\dot{E}$ (erg s$^{-1}$)} &
\colhead{$P_4/P$} & \colhead{Ref.} & \colhead{$L_b$ (erg s$^{-1}$)} & \colhead{Ref.} & \colhead{$L_b/{\dot E}$}  &
\colhead{$T_s$ (10$^{6}$ K)}
}
\startdata
B0943$+$10 &1.09 &3.49 &$1.0\times 10^{32}$ &$37.4^{+0.4}_{-1.4}$ &1 &$(5.0^{+0.6}_{-1.7})10^{28}$    &8&$(0.49^{+0.06}_{-0.16})10^{-3}$ &$3.1^{+0.9}_{-1.1}$ \\
B1113$+$16 &1.19 &3.73 &$8.8\times 10^{31}$ &$33\pm 3$            &2 &                                & &                               & \\
           &     &     &                    &$32\pm 4$            &3 &$(6.8^{+1.1}_{-1.3})10^{28}$    &9&$(0.77^{+0.13}_{-0.15})10^{-3}$ &$3.2^{+1.9}_{-1.0}$ \\
B0834$+$06 &1.27 &6.8  &$1.3\times 10^{32}$ &$30.2\pm 0.2$      &4 &$(8.6^{+14.2}_{-4.4})10^{28}$   &5&$(0.67^{+1.1}_{-0.6})10^{-3}$ &$2.0^{+2.0}_{-0.9}$\\
B1929$+$10 &0.23 &1.16 &$3.9\times 10^{33}$ &$50^{+15}_{-5}$      &5 &$(1.17^{+0.13}_{-0.4})10^{30}$ &10&$(0.29^{+0.04}_{-0.09})10^{-3}$ &$3.5^{+0.2}_{-0.5}$\\
B0656$+$14 &0.38 &55.0 &$3.8\times 10^{34}$ &$20\pm 1$            &6 &$(5.7^{+0.6}_{-0.8})10^{31}$    &11&$(1.5\pm 0.3)10^{-3}$ &$1.25^{+0.03}_{-0.03}$ \\
B1055$-$52 &0.19 &5.8 &$3.0\times 10^{34}$ & $22^{+11}_{-5}$&13&
$(1.6^{+0.88}_{-0.42})10^{31}$   &11&$(0.53^{+0.88}_{-0.42})10^{-3}$ &$1.8^{+0.06}_{-0.06}$ \\

B0628$-$28 &1.24 &7.12 &$1.5\times 10^{32}$ &$7\pm 1$             &6 &$(2.9^{+1.5}_{-0.8})10^{30}$    &12&$(1.9^{+1.0}_{-0.5})10^{-2}$ &$3.3^{+1.3}_{-0.6}$\\
B0826$-$34 &1.85 &0.99 &$6.2\times 10^{30}$ &$14\pm 1$            &7&$<1.45~10^{29}$                  &5&$<22~10^{-3}$ & \\

\enddata
\tablecomments{Errors in $L_b$ and $T_s$ correspond to 2$\sigma$ (90 \%
confidence) level. References: 1 - DR99; 2 - Paper I; 3 - HR07; 4 - RW07;
5 - this Paper; 6 - Paper II; 7 - G04; 8 - Z05; 9 - K06; 10 - M07; 11 -
DL05; 12 TO05; 13 B90 }
\end{deluxetable}

\clearpage

\begin{deluxetable}{ccccrrr}
\tablecaption{The \xmm\ EPIC observations of PSR B0826$-$34 and PSR B0834$+$06.}
\tablewidth{0pt}

\tablehead{
\multicolumn{2}{c}{Pointing direction} & \colhead{Sat.} & \colhead{Inst.\tablenotemark{a}} &
\colhead{Start time} & \colhead{End time} & \colhead{Exp.\tablenotemark{a}}
\\
\multicolumn{2}{c}{R.A. (J2000.0) Dec.} & \colhead{Rev.} & \colhead{} & \multicolumn{2}{l}{(UT)} &
\colhead{ks}
}
\startdata
  \multicolumn{7}{c}{PSR B0826$-$34 (Observation ID 0400020101):} \\
  08 28 16.6 & -34 17 07 & 1269  & PN & 2006-11-13 13:44:24 & 2006-11-14 09:19:30 & 38.83 \\
         &       &   & M1 &        13:22:03 &        09:19:35 & $-$   \\
         &       &   & M2 &        13:22:03 &        09:19:50 & $-$   \\
  \multicolumn{7}{c}{PSR B0834$+$06 (Observation ID 0501040101):} \\
  08 37 05.6 &  06 10 15 & 1454  & PN & 2007-11-17 13:44:24 & 2007-11-18 09:19:30 & 48.95 \\
         &       &   & M1 &        13:22:03 &        09:19:35 & 53.30 \\
         &       &   & M2 &        13:22:03 &        09:19:50 & 54.44 \\

\enddata

\tablenotetext{a}{The three EPIC instruments were operated in full frame CCD readout mode with 73 ms frame time
         for PN and 2.6 s for MOS with thin optical blocking filters.}
\tablenotetext{b}{Net exposure times after background screening.}
\label{tab-obs}
\end{deluxetable}

\clearpage
\begin{figure}
\includegraphics[scale=1]{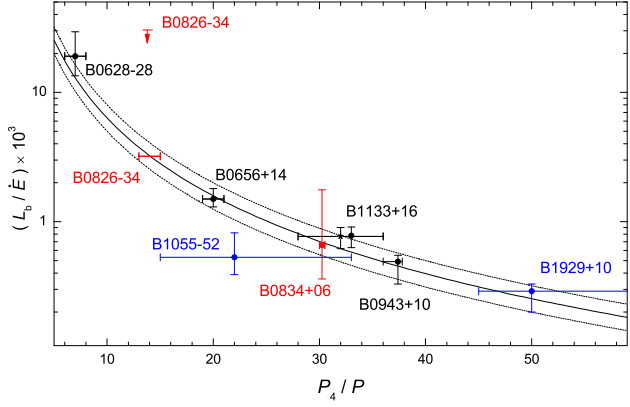}
\caption{The efficiency of thermal X-ray emission from a hot polar cap $L_b/\dot E$ versus circulation period $P_4$ of drifting subpulses in the
radio band. The solid curve represents the prediction of the PSG model (eq.~[1]), while the dotted curves correspond to uncertainties in determining
the moment of inertia (see Appendix A). The values of $P_4$ and $L_b$ along with their error bars (2$\sigma$) and references for the data are given
in Table 1.}
\end{figure}
\clearpage

\begin{figure}
\includegraphics[scale=1]{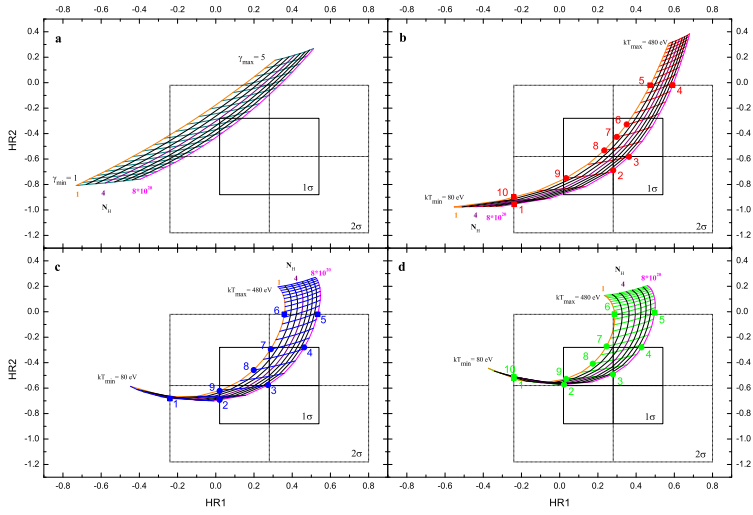} \caption{{\bf a.} Hardness ratios HR1 and HR2 derived for
           a grid of power-law model spectra with varying column density
       \nh\ and photon index $\gamma$ compared to the measured values from the EPIC-PN
       data of PSR B0834+06. The cross and box drawn with full lines indicate 1$\sigma$ and
       dotted lines 2$\sigma$ confidence regions.
       {\bf b.} As in Figure~2a but for an absorbed blackbody model with $kT$ ranging from 80 to 480 eV, with a step of 20 eV.
       {\bf c.} As in Figure~2b but for a model with blackbody and power-law component.
           Both components are absorbed by the same \nh\ and the relative (0.2$-$10.0 keV)
       flux ratio is 1:0.5, respectively.
       {\bf d.} As in Figure~2c but for a flux ratio of 1:1.}
\end{figure}
\clearpage
\begin{figure}
\includegraphics[scale=1]{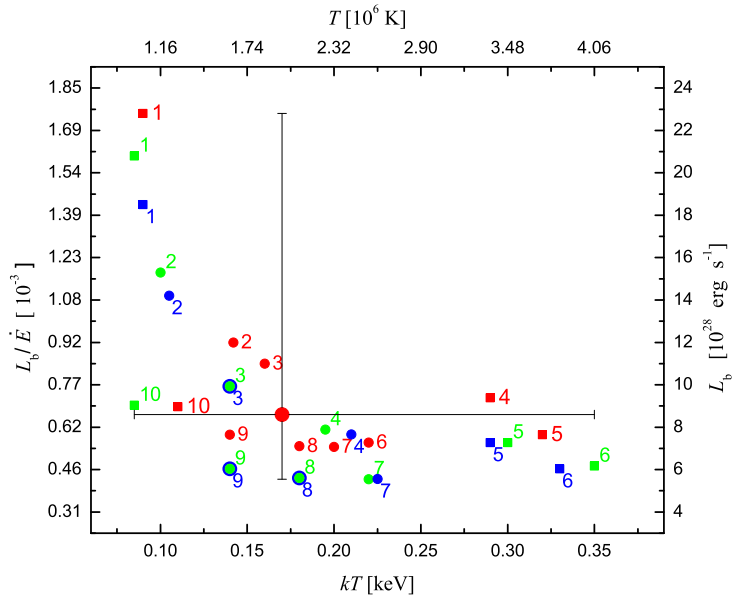}
\caption{Thermal luminosity $L_b$ and its efficiency $L_b/\dot E$ versus the polar cap temperature $kT$ for B0834+06 derived by means of XSPEC
spectral modelling for color marked and numbered points in Figures~2b-2d. The large red circle corresponds to the best fit
of the BB model and the error bars include the model uncertainties.}
\end{figure}

\end{document}